\documentclass[aps,prd,twocolumn]{revtex4-2}

\usepackage{graphicx}
\usepackage{amsmath}
\usepackage{amssymb}
\usepackage[table]{xcolor}
\usepackage[colorlinks,citecolor=blue,linkcolor=blue,urlcolor=blue,anchorcolor=blue]{hyperref}

\allowdisplaybreaks

\begin{document}

\title{Avoiding recollapse in an open-AdS universe via a self-tuning-like mechanism}

\author{Yupeng Zhang}
\author{Shuxun Tian}
\email{tshuxun@bnu.edu.cn}
\author{Zhengxiang Li}
\affiliation{School of Physics and Astronomy, Beijing Normal University, Beijing 100875, China}

\date{\today}

\begin{abstract}
We study whether an open FLRW universe with a negative cosmological constant can evade the eventual recollapse characteristic of an AdS-type universe. Within a power-law realization of Fab-Four theory, we solve the background equations numerically and analyze the asymptotic dynamics. For the representative branch and parameter choice studied here, we find that the scalar sector provides a self-tuning-like compensation for the negative $\Lambda$, while the curvature term remains unscreened. As a result, the universe can continue expanding instead of recollapsing. Instead, the universe evolves toward a curvature-dominated linear-expansion regime, $a\propto t$. To probe the underlying compensation mechanism, we further analyze an auxiliary zero-curvature subsystem using Poincar\'e compactification. In the \(\Lambda<0\) domain, there exist background trajectories that approach a critical point at infinity. Near this point, the compensating scalar-\(\Lambda\) sector becomes stiff-like, $w_{\phi+\Lambda}\to1$, so that the system effective energy density redshifts faster than curvature ($w_k=-1/3$). Although this auxiliary analysis does not cover the full curved cosmology, it is consistent with and qualitatively supports the numerical finding that the net $\phi+\Lambda$ contribution becomes subdominant to curvature, thereby preventing recollapse despite $\Lambda<0$. This extends the application of the self-tuning mechanism to the AdS region and offers a possibility for the AdS Universe predicted by string theory to become a reality.

\end{abstract}

\maketitle

\section{INTRODUCTION}

The cosmological constant problem remains one of the central open problems in modern cosmology and gravitational theory \cite{1989RvMP...61....1W,2001LRR.....4....1C}. Observationally, a positive cosmological constant provides the simplest effective description of the present accelerated expansion. However, the self-consistency at high energies appears to disfavor a single positive cosmological constant \cite{2018arXiv180608362O,2019PhLB..788..180O}. One possible approach is to replace the cosmological constant by a negative one, corresponding to an anti--de Sitter (AdS) vacuum. In particular, AdS vacua arise naturally in string theory and occupy a central place in the string landscape and holographic constructions \cite{1999IJTP...38.1113M,2006PhR...423...91G,2007RvMP...79..733D}. Nevertheless, a negative $\Lambda$ is typically accompanied by serious consequences, as it generally suppresses cosmic expansion and eventually drives the universe back to collapse. \citet{2003PhRvD..67h3501C} showed that, in a universe with a negative cosmological constant, the introduction of quintessence cannot alter the final fate of collapse within certain regions of parameter space. This naturally raises an important question: in the context of cosmological evolution, is there any mechanism that can keep an AdS universe expanding continuously and thereby change its ultimate fate?

An attractive approach to this question is the Fab-Four sector of Horndeski theory \cite{1974IJTP...10..363H,2011PThPh.126..511K,2012PhRvD..85j4040C}. It was originally identified as the unique subset of second-order scalar-tensor theories capable of realizing self-tuning of the cosmological constant on FLRW backgrounds, such that the vacuum energy can be dynamically screened, thereby offering a possible resolution to the cosmological constant problem \cite{2012PhRvL.108e1101C}. However, this self-tuning property is nontrivial in the presence of a negative $\Lambda$. For example, in \citet{2026arXiv260323263M}, Fig.~1 shows that no nontrivial critical point exists in the region with $\Lambda<0$, indicating that self-tuning does not operate there and the vacuum energy is not screened. This motivates us to explore whether Fab-Four dynamics can self-tune a negative \(\Lambda\) and thereby allow an open negative-\(\Lambda\) FLRW universe to avoid recollapse.

This paper is organized as follows. In Sec.~\ref{sec:MODEL} we introduce a power-law realization of Fab-Four and derive the field equations on an FLRW universe. In Sec.~\ref{sec:Analysis Of Evolution} we present the numerical evolution of an AdS universe in the presence and absence of $\phi$, respectively. Then, in Sec.~\ref{sec:Phase Diagram Analysis}, we analyze the asymptotic behavior of the system in an auxiliary zero-curvature subsystem using Poincar\'e compactification. The construction of the dynamical system and the choice of dimensionless variables are presented in Sec.~\ref{sec:Construction of Dynamical Systems}, while the corresponding phase portrait analysis and coordinate transformation are given in Sec.~\ref{sec:Phase Analysis}. Finally, in Sec.~\ref{sec:Conclusions}, we summarize our results and discuss the physical interpretation and limitations of this mechanism.

\section{MODEL}
\label{sec:MODEL}
The Fab-Four framework consists of four base Lagrangians, and its action can be written as 
\begin{align}
    S&=\int  \mathrm{d}^4x\sqrt{-g}[  {\cal L}_\mathrm{john} + {\cal L}_\mathrm{paul} + {\cal L}_\mathrm{george} + {\cal L}_\mathrm{ringo} - \frac{c^3 }{\kappa}\Lambda]  \nonumber \\
    &\quad+S_M,\label{Fun:action}
\end{align}
where $S_M$ denotes the action of matter and radiation. Each of these four components contains a scalar function $V_i(\phi)$. The four components can be written as follows \cite{2012PhRvL.108e1101C}:
\begin{align}
    {\cal L}_\mathrm{john}&= \frac{c^3 }{2 \kappa} V_{\rm john}(\phi)G^{\mu\nu}\nabla_{\mu}\phi\nabla_{\nu}\phi,\\
    {\cal L}_\mathrm{paul}&= \frac{c^3 }{2 \kappa} {{V_{\rm paul}(\phi)P^{\mu\nu\alpha\beta}\nabla_{\mu}\phi\nabla_{\alpha}\phi\nabla_{\nu}\nabla_{\beta}\phi}},\\
    {\cal L}_\mathrm{george}&= \frac{c^3 }{2 \kappa} V_{\rm george}(\phi)R,\\
    {\cal L}_\mathrm{ringo}&= \frac{c^3 }{2 \kappa} V_{\rm ringo}(\phi)\hat{G},
\end{align}
where $P^{\mu\nu\alpha\beta}=1/4\varepsilon^{\mu \nu \lambda \sigma}R_{\lambda\sigma\gamma\delta} \varepsilon^{\alpha \beta \gamma \delta}$, $\kappa=8 \pi G$, and $\hat{G}=R_{\mu \nu \alpha \beta }R^{\mu \nu \alpha \beta}-4 R_{\mu \nu}R^{\mu \nu }+R^2$ is the Gauss-Bonnet combination.

In analogy with Brans-Dicke theory \cite{1961PhRv..124..925B}, we take the scalar field $\phi$ to be dimensionful, whereas the parameters in $V_i(\phi)$ are assumed to be dimensionless and of order unity. 
In this paper, the dimension of $\phi$ is taken to be $[L^{-1}]$. In the original Fab-Four construction, the scalar functions \(V_i(\phi)\) are left arbitrary functions of the scalar field \cite{2012PhRvL.108e1101C}. We adopt a monomial power-law realization in this work. This choice is motivated by two considerations. First, power-law functions possess the property of scale self-similarity. Second, power-law potentials have long served as benchmark choices in studies of cosmic evolution, with representative examples appearing in chaotic inflation \cite{1983PhLB..129..177L} and tracker quintessence \cite{1988PhRvD..37.3406R}. Considering that the action has dimension $[S]=[ML^2T^{-1}]$, while the overall factor has dimension $\left[c^3/2\kappa\right]=[MT^{-1}]$, and taking $[\phi]=[L^{-1}]$,
the scalar functions \(V_i(\phi)\) are chosen as
\begin{align}
V_{\rm john}(\phi) &= \alpha_1 \phi^{-4}, \label{Fun:Vjohn}\\
V_{\rm paul}(\phi) &= \alpha_2 \phi^{-7}, \label{Fun:Vpaul}\\
V_{\rm george}(\phi) &= \alpha_3 \phi^{0}, \label{Fun:Vgeorge}\\
V_{\rm ringo}(\phi) &= \alpha_4 \phi^{-2}, \label{Fun:Vringo}
\end{align}
where $\alpha_i$ are dimensionless parameters. Since $\alpha_3$ in $V_{\rm george}$ can be absorbed into a rescaling of $\kappa$, we set $\alpha_3=1$. The model is then characterized by the three remaining parameters $\alpha_1$, $\alpha_2$, and $\alpha_4$. In the original Fab-Four construction, the self-tuning filter leaves four base Lagrangians, each multiplied by an arbitrary function of the scalar field. This functional freedom, however, should not be interpreted as implying that every assignment of the scalar functions gives a self-tuning branch. As shown in \citet{2012PhRvD..85j4040C}, the self-tuning condition excludes the special choice
\begin{equation}
\{V_{\rm john},V_{\rm paul},V_{\rm george}\}
=
\{0,0,\mathrm{constant}\}.
\end{equation}
The power-law realization adopted in Eq.~(\ref{Fun:Vjohn}) - (\ref{Fun:Vgeorge}) does not fall into this degenerate case.

By varying Eq.~(\ref{Fun:action}) with respect to the components of the metric tensor and its first derivatives, one obtains the field equations of the system. Meanwhile, varying it with respect to $\phi$ yields the equation of motion for the scalar field $\phi$. Under the FLRW metric, the temporal component and spatial component of the field equations, together with the equation of motion of the scalar field, can be written as follows:
\begin{align}
    H^2 +\frac{k c^2}{a^2}&=\frac{\Lambda c^2}{3} +\frac{8\pi G}{3}\rho + F_1(a,\phi) + k F_2(a,\phi) \label{Fun:F00},\\
    \frac{k c^2}{3a^2} + \frac{H^2}{3} + \frac{2\ddot{a}}{3a} &= \frac{\Lambda c^2}{3}-\frac{8 \pi G}{3 c^2} p + F_3(a,\phi) + k F_4(a,\phi),\label{Fun:F11}
\end{align}
\begin{align}
 &\alpha_1\left(  \frac{12 H^2 \dot{\phi}^2}{c^2 \phi^5} - \frac{6 H^3 \dot{\phi}}{c^2 \phi^4}  - \frac{12 H \ddot{a} \dot{\phi}}{c^2 a \phi^4} - \frac{6 H^2\ddot{\phi}}{c^2 \phi^4}\right) + \alpha_4\frac{48 H^2 \ddot{a}}{c^2 a \phi^3} \nonumber\\
 &+ \alpha_2  \left( \frac{42H \dot{\phi}^3}{c^4 \phi^8} - \frac{27 H^2 \ddot{a} \dot{\phi}^2}{c^4 a \phi^7} - \frac{18 H^3 \dot{\phi} \ddot{\phi}}{c^4 \phi^7}  \right)= \nonumber \\
 &\quad\alpha_1 k \left( - \frac{6  H\dot{\phi}}{a^2 \phi^4} + \frac{12  \dot{\phi}^2}{a^2 \phi^5} - \frac{6  \ddot{\phi}}{a^2 \phi^4}\right) + \alpha_4  k\frac{48  \ddot{a}}{a^3 \phi^3} \nonumber \\
 &\quad\quad+ \alpha_2 k \left( \frac{42  H \dot{\phi}^3}{c^2 a^2 \phi^8}  - \frac{9  \ddot{a} \dot{\phi}^2}{c^2 a^3 \phi^7} - \frac{18  H \dot{\phi} \ddot{\phi}}{c^2 a^2 \phi^7}  \right),\label{Fun:Motivation Function}
\end{align}
where
\begin{align}
    F_1(a,\phi)&=\alpha_1 \frac{3 H^2 \dot{\phi}^2}{2c^2 \phi^4}  - \alpha_2 \frac{5 H^3 \dot{\phi}^3}{2c^4 \phi^7} + \alpha_4 \frac{8H^3 \dot{\phi}}{c^2 \phi^3},\\
    F_2(a,\phi)&=\alpha_1\frac{ \dot{\phi}^2}{2 a^2 \phi^4}  - \alpha_2 \frac{3 H\dot{\phi}^3}{2c^2 a^2 \phi^7} + \alpha_4 \frac{8 H\dot{\phi}}{a^2 \phi^3},\\
    F_3(a,\phi)&=\alpha_1 \left(\frac{H^2 \dot{\phi}^2}{6 c^2 \phi^4} -  \frac{4 H \dot{\phi}^3}{3 c^2 \phi^5} + \frac{\ddot{a} \dot{\phi}^2}{3 c^2 a \phi^4} + \frac{2 H \dot{\phi} \ddot{\phi}}{3 c^2 \phi^4}\right) \nonumber\\
    &+ \alpha_2 \left( \frac{7 H^2 \dot{\phi}^4}{2 c^4 \phi^8} - \frac{H \ddot{a} \dot{\phi}^3}{c^4 a \phi^7} - \frac{3 H^2 \dot{\phi}^2 \ddot{\phi}}{2 c^4 \phi^7} \right)\nonumber\\
    &- \alpha_4 \left( \frac{8 H^2 \dot{\phi}^2}{c^2 \phi^4} + \frac{16 H \ddot{a} \dot{\phi}}{3 c^2 a \phi^3} + \frac{8 H^2 \ddot{\phi}}{3 c^2 \phi^3}\right),\\
    F_4(a,\phi)&=- \alpha_1 \frac{ \dot{\phi}^2}{6 a^2 \phi^4}
    + \alpha_2 \left(\frac{7  \dot{\phi}^4}{6 c^2 a^2 \phi^8} - \frac{\dot{\phi}^2 \ddot{\phi}}{2 c^2 a^2 \phi^7}\right) \nonumber\\
    & - \alpha_4 \left(\frac{8 \dot{\phi}^2}{a^2 \phi^4}  + \frac{8 \ddot{\phi}}{3 a^2 \phi^3}\right).
\end{align}
This theory belongs to the Horndeski framework, so that even though second-order derivatives of $\phi$ appear in the Lagrangian, the equations of motion remain second order equations. In the above equations, the effect of the scalar field $\phi$ is encoded in $F_i(a,\phi)$. If these four functions all vanish, the system reduces to the standard cosmological framework, but the value of $G$ may be modified. In the following, based on Eq.~(\ref{Fun:F00}) and for given initial conditions, we perform a numerical evolution to investigate how an AdS universe evolves with and without the scalar field, respectively.

\subsection{ Analysis of cosmological evolution}
\label{sec:Analysis Of Evolution}
\begin{figure}[htbp]
\includegraphics[width=\columnwidth]{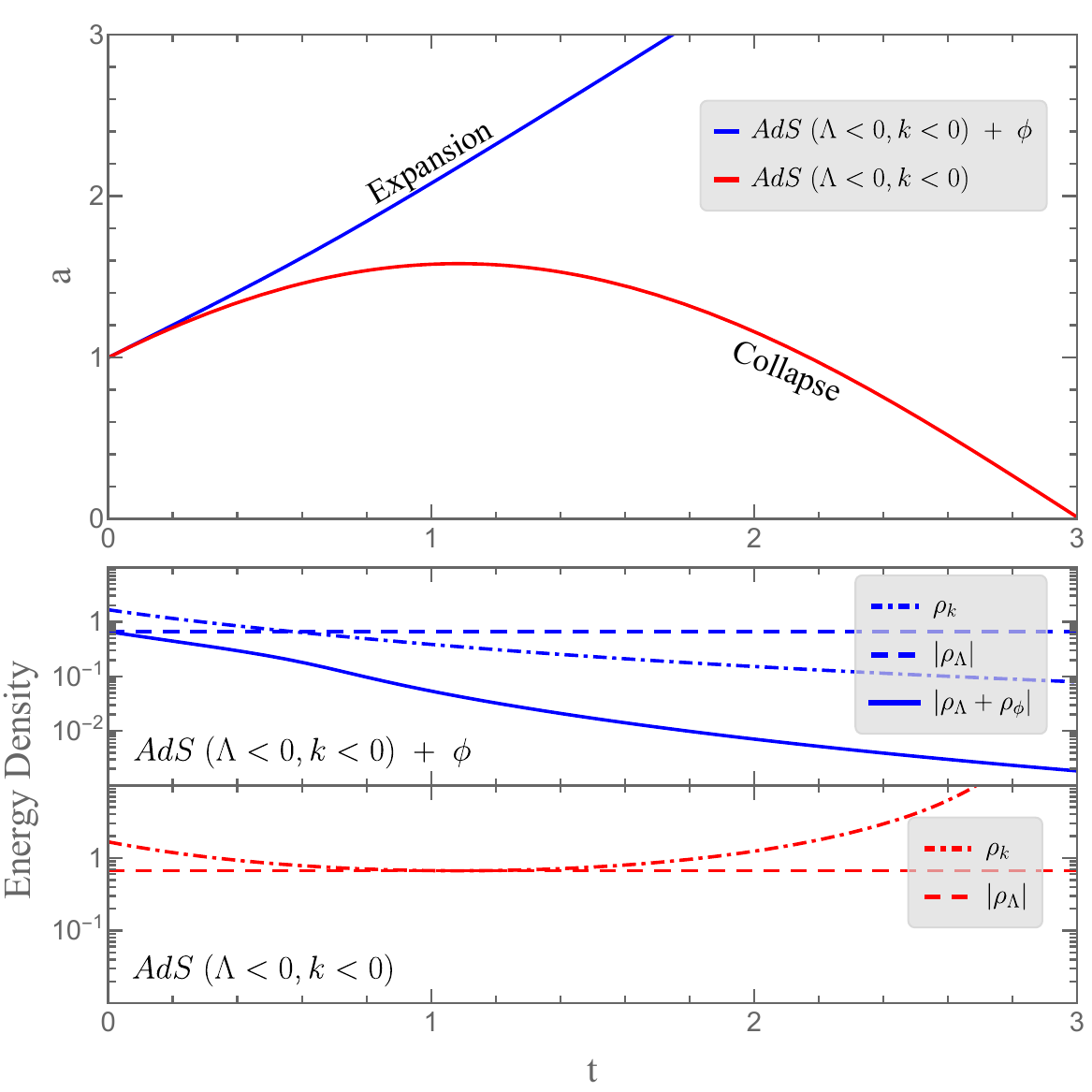}
\caption{\label{fig:AdS evolution} Evolution of the AdS vacuum with and without $\phi$. In the upper panel, the horizontal axis denotes cosmic time, while the vertical axis represents the scale factor. The red curve corresponds to the evolution of the AdS vacuum, whereas the blue curve shows the evolution with the scalar field $\phi$ included. The lower panel displays the evolution of the energy densities of different components. Since a negative $\Lambda$ contributes a negative energy density, absolute values are taken for some quantities for the purpose of clearer visualization, as indicated in the legend. We numerically obtained the above results using Eqs.~(\ref{Fun:F11}) and (\ref{Fun:Motivation Function}). The initial conditions for AdS with $\phi$ are $p=0,\;\phi_0=1,\; \dot{\phi}_0=-0.001,\; a_0=1,\; \dot{a}_0=1,\;\alpha_i=1,\; k=-5/3,\; \Lambda=-2.016$. For the case without $\phi$, the initial conditions are $p=0,\; a_0=1,\; \dot{a}_0=1,\;\alpha_i=1,\; k=-5/3,\; \Lambda=-2$. The speed of light and gravitational constant are set $G=1,\ c=1$.}
\end{figure}

The AdS vacuum is characterized by negative curvature and a negative $\Lambda$, and contains neither matter nor radiation. We set the initial curvature energy density $\rho_k$ and the cosmological-constant energy density $\rho_\Lambda$ are expected to be of the same order of magnitude. As a representative numerical example, we take the remaining dimensionless parameters to be $\alpha_i=1$ and specify the initial conditions as given in the caption of Fig.~\ref{fig:AdS evolution}. We then compare the evolution of the open-AdS background in the presence and absence of the scalar field. 

The upper panel shows that a pure AdS vacuum cannot sustain continuous expansion, since the negative $\Lambda$ eventually drives the universe back to collapse. By contrast, after the scalar field is introduced, the scale factor $a$ grows approximately linearly with time, which is similar to that in the $R_h=ct$ universe \cite{2012MNRAS.419.2579M}, indicating that the spacetime evolution enters a curvature-dominated regime. The lower panel presents the evolution of the energy densities of different components, where the quantities associated with $\Lambda$ are shown in terms of their absolute values for clarity. One can see that $\rho_\Lambda$ remains constant throughout the evolution, which is consistent with the nature of the cosmological constant. Meanwhile, $|\rho_\Lambda+\rho_\phi|$ continuously decreases with time, and by $t=3$ it has dropped from the same order of magnitude as $\rho_k$ to about two orders of magnitude smaller than $\rho_k$. We also calculate that $w_{\phi+\Lambda}\to 1/3$ in this case, which further indicates that it decays faster than the curvature term ($w_k=-1/3$). This indicates that the scalar field dynamically compensates for the negative $\Lambda$, but does not simultaneously cancel the curvature component. This behavior differs from the result in \citet{2026arXiv260323263M}, where both the curvature term and the cosmological-constant term are canceled simultaneously, suggesting that the dimension assigned to $\phi$ and the specific choice of the scalar functions can affect the tracking behavior of self-tuning. This confirms that self-tuning a negative $\Lambda$ is nontrivial in Fab-Four theory.

\section{Dynamical-system analysis}
\label{sec:Phase Diagram Analysis}
To investigate the global background evolutionary structure of this scenario in greater detail, we employ the dynamical-system approach to analyze evolution. For a general introduction, see \citet{2006IJMPD..15.1753C,2010deto.book.....A} and \citet{,2018PhR...775....1B}. As shown above, the present theory allows a universe with negative $\Lambda$ to undergo sustained expansion. We set $k=0$ and neglect the contributions from matter and radiation in the following analysis. This reduces the dynamical system to two dimensions, which facilitates the subsequent study of the asymptotic dilution rate of scalar field with $\Lambda$ and allows the background flow structure to be illustrated through the phase portrait. Since the curvature term has been neglected, the following analysis is not directly identical to the results shown in Fig.~\ref{fig:AdS evolution}. 
\subsection{Construction of the autonomous dynamical system}
\label{sec:Construction of Dynamical Systems}

To make the final expressions dimensionless and to keep $\dot{\phi}/\phi$ of the same order as $H$ at infinity, we choose the following dimensionless variables:
\begin{align}
    x_1\equiv \frac{\dot{\phi}}{c \phi^2},\;x_2\equiv \frac{H}{c \phi}\label{Fun:dimensionless variable}.
\end{align}
Using Eqs.~(\ref{Fun:F00}), (\ref{Fun:F11}), and (\ref{Fun:Motivation Function}), the evolution equations can be written as autonomous equations:
\begin{align}
   \frac{\mathrm{d}x_1}{\mathrm{d}N}&=P(x_1,x_2) = x_2 \gamma_2 - \frac{2 x_1{}^2}{x_2},\quad \label{Fun:differential equation P}\\
   \frac{\mathrm{d}x_2}{\mathrm{d}N}&= Q(x_1,x_2) =x_2 \gamma_1 -x_1,\label{Fun:differential equation Q}
\end{align}
Where $N=\ln a$ is the e-folding number, and $\gamma_i$ denote the quantities obtained by eliminating the second-order derivatives by combining Eqs.~(\ref{Fun:F11}) and (\ref{Fun:Motivation Function}), which can be written as follows:
\begin{align}
    \gamma_1&=\frac{\dot{H}}{H^2}=\frac{\ddot{a}}{a H^2}-1=\frac{X_1}{X_3},\\
    \gamma_2
    &=\frac{\ddot{\phi}}{H^2 \phi}=\frac{X_2}{X_3},
\end{align}
where,
\begin{align}
     X_1&=-4\alpha_1 (9 \alpha_1 + 8 \alpha_4)x_1{}^2 -2\alpha_1 \alpha_2 x_1{}^4 \nonumber\\
     &+[-192 \alpha_1 \alpha_4 x_1 + 16 \alpha_2 (9 \alpha_1 + 4 \alpha_4) x_1{}^3] x_2\nonumber\\
     &\quad+ (-256 \alpha_4{}^4 + 384 \alpha_2 \alpha_4 x_1{}^2 - 135 \alpha_2{}^2 x_1{}^4) x_2{}^2, \label{Fun:B1numerator}\\
     X_2&=16\alpha_1 x_1{}^2 + 24 \alpha_1{}^2 x_1{}^4 -24 \alpha_1x_1 x_2 + 36\alpha_1{}^2 x_1{}^3 x_2\nonumber\\
     &- 56 \alpha_2 x_1{}^3 x_2 +192 \alpha_1 \alpha_4 x_1{}^3 x_2 -104 \alpha_1 \alpha_2 x_1{}^5 x_2 \nonumber\\
     &\;+ [-64 \alpha_4 + 12 (3 \alpha_2 +32 \alpha_1\alpha_4 + 64 \alpha_4{}^2) x_1{}^2] x_2{}^2 \nonumber\\
     &\;\;+ [-16\alpha_2 (9 \alpha_1 + 20 \alpha_4) x_1{}^4 + 105 \alpha_2{}^2 x_1{}^6] x_2{}^2 \nonumber\\
     &\;\;\;+ (768 \alpha_4{}^2 x_1 - 672 \alpha_2 \alpha_4 x_1{}^3 + 135 \alpha_2{}^2 x_1{}^5) x_2{}^3, \label{Fun:B2numerator}\\
     X_3&=12 \alpha_1{}^2 x_1{}^2 -8(3 \alpha_2 x_1 -8 \alpha_1 \alpha_4 x_1-6 \alpha_1 \alpha_2 x_1{}^3) x_2 \nonumber \\
     &8 \alpha_1 + (256 \alpha_4{}^2  -96 \alpha_2 \alpha_4 x_1{}^2  + 45 \alpha_2{}^2 x_1{}^4 )x_2{}^2.
\end{align}
After introducing the dimensionless variables, the field equation Eq.~(\ref{Fun:F00}) takes the form
\begin{align}
    1=\frac{3}{2}\alpha_1 x_1{}^2 + 8\alpha_4 x_1 x_2 - \frac{5}{2} \alpha_2 x_1{}^3 x_2 + \frac{\Lambda c^2}{3 H^2}.
\end{align}
In this work, we focus on the case with $\Lambda<0$, which provides a boundary for the evolution of $x_1$ and $x_2$, namely,
\begin{align}
    \frac{3}{2}\alpha_1 x_1{}^2 + 8\alpha_4 x_1 x_2 - \frac{5}{2} \alpha_2 x_1{}^3 x_2>1.
\end{align}

It should be noted that the autonomous system above is formulated with respect to the e-folding number
$N=\ln a$, rather than the cosmic time $t$. Since $dN/dt=H$, the arrows in the phase portrait indicate the direction of increasing scale factor. Therefore, on an expanding branch with $H>0$, they coincide with the forward direction of cosmic time, whereas on a collapsing branch with $H<0$, this correspondence is reversed. In the present work we restrict the phase-space discussion to the expanding branch relevant to the avoidance of recollapse. We therefore do not attempt to infer the time evolution of collapsing solutions directly from the phase portrait, and leave such branches outside the scope of the present analysis.

Moreover, Eqs.~(\ref{Fun:differential equation P}) and (\ref{Fun:differential equation Q}) is symmetric under the central inversion. Under the transformation $(x_1,x_2)\mapsto(-x_1,-x_2)$, one finds
\begin{align}
    P(-x_1,-x_2)&=-P(x_1,x_2),\\
    Q(-x_1,-x_2)&=-Q(x_1,x_2).
\end{align}
It follows that, whenever $x(N)$ is a solution, $-x(N)$ is also a solution with respect to the same e-folding number $N$. Hence the phase flow is equivariant under the central inversion $(x_1,x_2)\mapsto (-x_1,-x_2)$, and the oriented phase portrait is centrally symmetric about the origin.

\subsection{Global phase portrait and asymptotic dynamics}
\label{sec:Phase Analysis}
In the Fab-Four construction, self-tuning does not require all four fundamental Lagrangians to be active simultaneously. For example the ${\cal L}_\mathrm{george}$ alone can realize self-tuning when \(V_{{\rm george}}(\phi)\neq \mathrm{const}\) \cite{2012PhRvL.108e1101C}. 
In this case, the reduced theory is equality to Brans-Dicke gravity with \(\omega_{\rm BD}=0\), which is excluded by Solar System tests \cite{2014LRR....17....4W} in the absence of an additional screening mechanism. As for ${\cal L}_\mathrm{ringo}$, it cannot realize self-tuning on its own without the support of ${\cal L}_\mathrm{john}$ or ${\cal L}_\mathrm{paul}$ \cite{2012PhRvL.108e1101C}. We therefore prefer to construct the system with all four Lagrangians included simultaneously, and do not further consider the marginal regions of the parameter space. For the positive parameter choices tested, no qualitative change was found in the evolution. It should be emphasized that the phase portrait discussed below is a background-level dynamical phase portrait. It characterizes the global structure of background trajectories, but should not be interpreted as the perturbatively viable region of the phase space.

\begin{figure*}
\centering
\includegraphics[width=\textwidth]{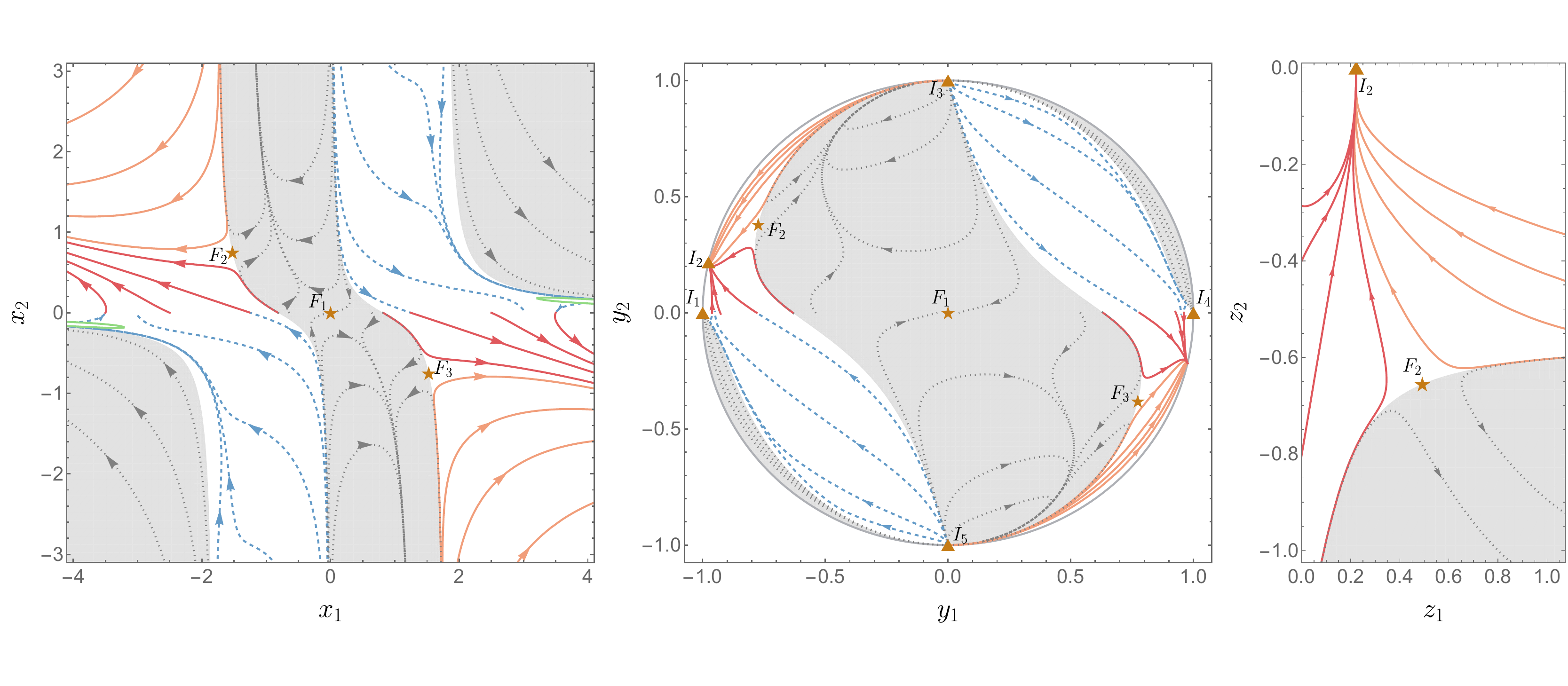}
\caption{\label{fig:Phase Diagram} Phase portrait of the dynamical system under different projections, for $\alpha_1=\alpha_2=\alpha_4=1$. The left panel shows the phase portrait on the $x_1$--$x_2$ plane, the middle panel shows the projection of the Poincar\'e sphere along the polar direction, and the right panel illustrates the asymptotic behavior as $x_2\to\infty$. The curves denote trajectories with different initial conditions, and the arrows indicate the direction of increasing scale factor; the same initial conditions are used in all panels. The white and gray regions correspond to $\Lambda<0$ and $\Lambda>0$, respectively. Stars mark the finite critical points, while triangles mark the critical points at infinity. The green solid line indicates the singularity of the system, which is omitted in the other panels.}
\end{figure*}

The phase portrait of the dynamical system for the choice $\alpha_i=1$ is shown in the left panel of Fig.~\ref{fig:Phase Diagram}.
The white region in the figure corresponds to the parameter space with $\Lambda<0$. Within the finite part of this parameter space, there exists only one critical point, located on the boundary. Because Eqs.~(\ref{Fun:B1numerator}) and (\ref{Fun:B2numerator}) contain high-order terms, their analytic expressions are highly complicated. We therefore calculate only the numerical Jacobian matrix, which shows that it is a saddle point. Since $x_2$ is inversely proportional to $\phi$ in the dimensionless variables, crossing the $x_1$ axis cannot, strictly speaking, be achieved through a finite variation of the field value. For this reason, the trajectories that cross the $x_1$ axis in the phase portrait are shown separately as red solid and blue dashed lines. Because of the central symmetry of the system, we then focus on the dynamical behavior in the second quadrant. Since there is only one saddle point in the finite region, this suggests that the trajectories in the second quadrant may generally evolve toward negative infinity. To examine this behavior, we calculate the evolution of $w_{\phi+\Lambda}$ as a function of the e-folding number for the sum of the scalar-field component and the negative cosmological-constant component along the dynamical trajectories, and find that $w_{\phi+\Lambda} \to 1$. 
This indicates that, as the system evolves, the scalar-field component rapidly decreases to a magnitude comparable to that of the cosmological-constant component. Combined with the result shown in Fig.~\ref{fig:AdS evolution}, this tracking behavior appears in both cases in which the initial scalar-field value is larger or smaller than $|\Lambda|$. 

To study the global phase portrait structure and, in particular, the asymptotic behavior at infinity, we perform a Poincar\'e compactification, a standard tool in global dynamical-systems analyses that has also been applied in cosmological phase portrait studies \cite{Tian2026}. Under this transformation, the points at infinity in the original phase portrait are mapped onto the equator of the Poincar\'e sphere, which allows the critical points at infinity to be identified and classified systematically. The transformation is given by:
\begin{align}
    y_1&=\frac{x_1}{\sqrt{1+x_1{}^2+x_2{}^2}},\\
    y_2&=\frac{x_2}{\sqrt{1+x_1{}^2+x_2{}^2}},\\
    y_3&=\frac{1}{\sqrt{1+x_1{}^2+x_2{}^2}}.
\end{align}
This transformation can be understood as placing a sphere outside the original phase portrait and mapping each point in the phase portrait to the intersection between the sphere and the line connecting that point to the center of the sphere. According to Eqs.~(\ref{Fun:B1numerator}) and (\ref{Fun:B2numerator}), one can then derive the autonomous differential equations $\mathrm{d}y_1/\mathrm{d}N$ and $\mathrm{d}y_2/\mathrm{d}N$ under this projection. Since the Poincar\'e compactification does not destroy the symmetry of the original phase portrait, it is still sufficient to restrict the discussion to the upper half of the phase portrait. The projection of the Poincar\'e sphere along the $y_3$ axis is shown in the middle panel of Fig.~\ref{fig:Phase Diagram}. Owing to the centrosymmetry of the system, there are three critical points at infinity on the equator of the Poincar\'e sphere, denoted by $I_1$, $I_2$, and $I_3$. The trajectories corresponding to the red and orange solid curves are also shown in the compactified phase portrait, and both are seen to evolve toward $I_2$. To determine the location and background dynamical stability of this point, we introduce the following projected coordinates:
\begin{eqnarray}
    z_1=-\frac{x_2}{x_1},\;z_2=\frac{1}{x_1}.
\end{eqnarray}
The essence of this coordinate choice is to map the region at infinity along the $x_1$ direction onto finite positions on the horizontal axis, so that the critical points originally located at infinity can be represented at finite coordinates. Likewise, under this projection, the autonomous system can be written as:
\begin{align}
    \frac{\mathrm{d}z_1}{\mathrm{d}N}&= -z_1 z_2 P\left(\frac{1}{z_2},-\frac{z_1}{z_2}\right) - z_2 Q\left(\frac{1}{z_2},-\frac{z_1}{z_2}\right) \label{Fun:differential equation z_1 in projection} ,\\
    \frac{\mathrm{d}z_2}{\mathrm{d}N}&= - z_2{}^2 P\left(\frac{1}{z_2},-\frac{z_1}{z_2}\right).\label{Fun:differential equation z_2 in projection}
\end{align}
It should be noted that the highest powers appearing in $X_1$, $X_2$, and $X_3$ are all of sixth order. As a result, the highest power in the numerator of $P$ and $Q$ is of seventh order, whereas that in the denominator is sixth order. Therefore, no additional time rescaling is required to ensure that the autonomous differential equations remain finite in the vicinity of $z_2=0$. The corresponding phase portrait is shown in the right panel of Fig.~\ref{fig:Phase Diagram}. To determine the location of the critical points at infinity, we take the limit $z_2\to 0$ in Eqs.~(\ref{Fun:differential equation z_1 in projection}) and (\ref{Fun:differential equation z_2 in projection}). In this limit, the equations are dominated by the highest-order terms in $P$ and $Q$, and the corresponding autonomous system can be written as
\begin{align}
    \lim_{z_2\to 0}\frac{\mathrm{d}z_1}{\mathrm{d}N}&= - \frac{-15 \alpha_2{}^2 z_1{}^2 + 135 \alpha_2{}^2 z_1{}^3}{45 \alpha_2{}^2 z_1{}^2} \nonumber\\
    &\quad\quad- \frac{-45 \alpha_2{}^2 z_1{}^2 + 135 \alpha_2{}^2 z_1{}^3}{45 \alpha_2{}^2 z_1{}^2}\nonumber\\
    &=\frac{4}{3}- 6 z_1\label{Fun:critical point z_1},\\
    \lim_{z_2\to 0}\frac{\mathrm{d}z_2}{\mathrm{d}N}&=0.
\end{align}
The critical point $I_2$ is found to be located at $(2/9,0)$ in the projected coordinate system. This result is independent of the specific choice of the Lagrangian coefficients $\alpha_i$. However, as can be seen from Eq.~(\ref{Fun:critical point z_1}), the highest-order terms contain only contributions associated with the coefficient $\alpha_2$, indicating that ${\cal L}_\mathrm{paul}$ plays the dominant role in this asymptotic regime. This is also consistent with the fact that ${\cal L}_\mathrm{paul}$ carries the smallest power index of the scalar field $\phi$ among the four Lagrangians. Although the phase portrait appears to suggest an attractor-like behavior near $I_2$, the Jacobian analysis shows that this point is in fact a hyperbolic critical point, with eigenvalues $-6$ and $-3/2$, respectively. Therefore, this point is an attractor, and no center manifold is associated with it. We also find that $w_{\phi+\Lambda}\to1$ at infinity, indicating that the scalar field compensates for a negative $\Lambda$.

In addition, there is also a region with $\Lambda<0$ in the first quadrant, whose trajectories are shown by the blue dashed curves. Owing to the centrosymmetry of the system, their mirror images in the third quadrant connect to the red curve. This connection arises from the choice of dimensionless variables: according to Eq.~(\ref{Fun:dimensionless variable}), $x_2$ is inversely proportional to the scalar field, so crossing zero corresponds to $\phi$ jumping from infinity on one side to infinity on the other, which is not physically acceptable. Moreover, such a divergence of $\phi$ may also be in tension with the swampland criterion \cite{2018JHEP...08..143G,2019ForPh..6700037P}, and we therefore do not consider these trajectories further. In addition, since $\gamma_1$ and $\gamma_2$ are obtained by eliminating variables through the field equations, the system contains singularities associated with vanishing denominators, whose locations are marked by the green solid line in the upper-left panel of Fig.~\ref{fig:Phase Diagram}.

\section{Conclusions}
\label{sec:Conclusions}

In this work, we investigated a power-law realization of the Fab-Four theory and studied its background cosmological evolution on the open-FLRW branch. We found that, in contrast to the pure negative-$\Lambda$ case where the universe is driven toward recollapse, the inclusion of the scalar sector can qualitatively modify the evolution and allow the background to enter a sustained expanding regime. For the branch and parameter choice considered here, the scalar sector provides a self-tuning-like background compensation of the negative $\Lambda$, while leaving the open-curvature contribution unscreened. As a result, the net $\phi+\Lambda$ contribution can become dynamically subdominant to the curvature term at late times, giving a proof-of-principle example of a negative-$\Lambda$ Fab-Four background that avoids recollapse.

To clarify the mechanism behind this behavior, we reformulated the cosmological equations as an autonomous dynamical system and analyzed the corresponding phase-portrait structure in an auxiliary zero-curvature subsystem. This analysis shows that there exist background trajectories approaching a critical point at infinity, where the compensated scalar-$\Lambda$ sector becomes stiff-like, with $w_{\phi+\Lambda}\to 1$. The numerical evolution of the full open system and the auxiliary zero-curvature analysis are consistent with the same interpretation: the net scalar-$\Lambda$ component redshifts faster than the open-curvature component, whose effective equation of state is $w_k=-1/3$. Therefore, once the curvature term is restored, the late-time evolution is driven toward a Milne-like, curvature-dominated expanding regime rather than a recollapsing state.

The present model should be regarded as a proof-of-principle construction rather than a complete description of the observed Universe. In the realization considered here, the scalar field is introduced only to prevent the recollapse caused by a negative cosmological constant. It is not intended to account for inflation or late-time accelerated expansion. These additional ingredients may be incorporated in more general Horndeski self-adjustment scenarios \cite{2012AdAst2012E..50B,2012JCAP...12..026C,2022JCAP...10..075K}, but they lie beyond the scope of the present work.

A natural direction for future work is to examine the stability of the branch studied in this paper. Such an analysis is essential for gravitational theory, but it has not been carried out in the present work. Existing results provide useful reference points: the perturbative stability conditions have been formulated for general Horndeski cosmologies on flat FLRW backgrounds \cite{2012JCAP...02..007D}, and related stability analyses have been performed for specific Fab-Four branches, including the cosmological Fab-Four solutions and the John/George sector \cite{2012JCAP...12..026C,2012AdAst2012E..50B}. However, in the curved open-AdS evolution shown in Fig.~\ref{fig:AdS evolution}, the curvature term remains unscreened and becomes dynamically important. Therefore, the curved open-AdS background is not covered by the flat-FLRW assumptions of those analyses. Accordingly, the existing flat-FLRW stability results cannot be directly applied to the curved evolution shown in Fig.~\ref{fig:AdS evolution}. They may be applied to the zero-curvature subsystem used in the phase-portrait analysis shown in Fig.~\ref{fig:Phase Diagram}. However, such an analysis would not directly validate or rule out the theory. A complete stability analysis including curvature is therefore left to future work.

\section*{Acknowledgements}
This work was supported by the National Natural Science Foundation of China under Grant Nos.~12405050 and 12433001, and the Fundamental Research Funds for the Central Universities.

The authors acknowledge the use of ChatGPT (OpenAI, GPT-5.5 Thinking; accessed in June 2026) during the revision of this manuscript solely for language editing. No AI tool was used to generate scientific content or produce figures.

\section*{Data availability}
The data that support the findings of this article are not publicly available. The data are available from the authors upon reasonable request.

\bibliography{aps}

\end{document}